\documentclass[print]{aa}

\usepackage{graphicx}

\usepackage{natbib}
\usepackage{txfonts}

\begin{document}

\title {Effects of the galactic winds on the stellar metallicity distribution of dwarf spheroidal galaxies}
\titlerunning{Stellar metallicity distributions}

\author{Gustavo A. Lanfranchi\inst{1, 2, 3}
\and    Francesca  Matteucci\inst{3, 4}}

\institute{N\'ucleo de Astrof\'\i sica Te\'orica, CETEC, Universidade
Cruzeiro do Sul, R. Galv\~ao Bueno 868, Liberdade,  01506-000, S\~ao Paulo, SP, Brazil
\and IAG-USP, R. do Mat\~ao 1226, Cidade Universit\'aria, 
05508-900 S\~ao Paulo, SP, Brazil
\and Dipartimento di Astronomia-Universit\'a di Trieste, Via G. B. 
Tiepolo 11, 34131 Trieste, Italy
\and  I.N.A.F. Osservatorio Astronomico di Trieste, via G.B. Tiepolo 
11, I-34131}

\date{Received xxxx/ Accepted xxxx}

\abstract{}
{To study the effects of galactic winds on the stellar metallicity distributions and on the evolution of Draco and Ursa Minor dwarf spheroidal galaxies, we compared the predictions of chemical evolution models, adopting different prescriptions for the galactic winds, with the photometrically-derived stellar metallicity distributions (SMDs) of both galaxies.}
{The chemical evolution models for Draco and Ursa Minor, which are able to reproduce several observational features of these two galaxies, take up-to-date nucleosynthesis into account for intermediate-mass stars and supernovae of both types, as well as the effect of these objects on the energetics of the systems.}
{For both galaxies, the model that best fits the data contains an intense continuous galactic wind, occurring at a rate proportional to the star formation rate. Models with a wind rate assumed to be proportional only to the supernova rate also  reproduce the observed SMD, but do not match the gas mass, whereas the models with no galactic winds fail to reproduce the observed SMDs. In the case of Ursa Minor, the same model as in previous works reproduces the observed distribution very well with no need to modify the main parameters of the model ( $\nu$ = 0.1 Gyr$^{-1}$ and $w_i$ = 10).
In the case Draco, on the other hand, the observed SMD requires a model with a lower supernova type Ia thermalization efficiency ($\eta_{SNeIa}$ = 0.5 instead of $\eta_{SNeIa}$ = 1.0) in order to delay the galactic wind, whereas all the other parameters are kept the same ($\nu$ = 0.05 Gyr$^{-1}$, $w_i$ = 4).}
{The model results strongly suggest that intense and continuous galactic winds play a very important role in the evolution of local dSphs by removing a large fraction of the gas content of the galaxy and defining the pattern of the abundance ratios and of the SMD.
}

\keywords {stars: metallicity --  galaxies: Local Group -- galaxies: evolution -- galaxies: dwarf -- }

\maketitle

\section{Introduction}

The dwarf spheroidal galaxies (dSphs) of the Local Group, when discovered, were believed to be very simple systems, similar in some aspects to globular clusters. In fact, these galaxies are small systems ($R_T = \sim 2\; to \sim 6 \;kpc$) with low luminosities ($M_V \ge$ -14 mag, $\mu_V \sim-22$ mag.arcsec$^{-2}$), low metallicities ([Fe/H] $\sim$ -2.5 to -1.0 dex), low masses ($\sim 10^7 M_\odot$), and they are almost totally depleted of neutral gas (for detailed information see the reviews from Mateo 1998; Kunth $\&$ Ostlin 2000 and references therein). In spite of their simple appearance, the dSphs are very interesting systems with a complex evolution that has not yet been fully understood. Even though the general characteristics of these galaxies are similar, their individual evolutionary history might be quite different, as indicated by the star formation histories proposed for several of these systems (Dolphin et al. 2005) and by the abundance patterns observed in each galaxy (Bonifacio et al. 2000; Shetrone, Cot\'e, Sargent 2001; Shetrone et al. 2003; Tolstoy et al. 2003;
Bonifacio et al. 2004; Venn et al. 2004; Sadakane et al. 2004; Geisler et al. 2005;
Monaco et al. 2005) . Besides that, the mechanisms that halted the star formation (SF) and removed the gas content from these galaxies remain uncertain, even though several scenarios have been proposed (see Grebel et al. 2003 for a more detailed discussion). 

Both the evolution of local dSphs and the mechanism through which the gas was removed from these galaxies can be analysed by comparisons between detailed chemical evolution models and observed SMDs. In the past few years, several studies (both spectroscopic and photometric) have addressed the question of the SMDs in local dSphs (Bellazzini et al. 2002; Koch et al. 2006a,b; Helmi et al. 2006; Battaglia et al. 2006). These studies revealed that, although similar, the SMDs of each galaxy exhibit small differences reflecting different details in their evolution. The global picture, however, is the same for all galaxies and it is characterised by an apparent lack of metal-poor stars, a peak at low metallicities ([Fe/H] $\sim$ -1.6 dex), and a sharp decrease at the high-metallicity tail (Helmi et al. 2006). All these features are related to the evolution of the systems and also to the mechanism that removed the gas out of the galaxy and halted the SF. The slower a galaxy evolves (low SFR), the lower the number of metal-rich stars formed, and the lower the value of [Fe/H] corresponding to the peak in the metallicity distribution. Besides that, if a large fraction of the gas content of the galaxy is removed, the SF decreases substantially (even stops), and no star would be formed with  metallicities higher than the one of the gas at the moment of its removal. That would cause a sharp decrease in the high-metallicity tail, which would be more intense as the gas is removed faster.

In our chemical evolution models, the evolution of the galaxy and the loss of material are controlled mainly by the star formation rate (SFR) and by the occurrence of galactic winds. The SFR is specified by the efficiency of the SF, 
$\nu$ (the inverse of the SF timescale given in $Gyr^{-1}$), whereas the rate of the galactic winds is represented by the efficiency of the galactic wind, 
$w_i$, which is the ratio between the wind rate and the SFR (see sect. 3 for more details). Almost all models of chemical evolution proposed in the literature agree in suggesting that the abundance ratios observed in local dSphs are reproduced with a low SFR regime (roughly 10 times lower than in Blue Compact Galaxies and down to 100 times lower than the one in the solar neighbourhood) (Ikuta $\&$ Arimoto 2001; Carigi et al. 2002; Lanfranchi $\&$ Matteucci 2003, 2004; Fenner et al. 2006), but do not reach a consensus regarding the mechanism responsible for removing the gas.

While some models suggest that small systems, with masses in the range $10^6 - 10^8 M_\odot$, could have their gas content (especially the fresh metals injected into the interstellar medium) removed by galactic winds (Robertson et al. 2005; Kawata et al. 2006), others argue that only external factors could remove a large fraction of the interstellar medium (ISM) of the dSphs (Ferrara $\&$ Tolstoy 2000; Marcolini et al. 2006). Lanfranchi $\&$ Matteucci (2003, 2004 - LM03, LM04) and Lanfranchi, Matteucci $\&$ Cescutti 2006a (LMC06a), in particular, were able to reproduce several abundance ratios, as well as the present-day gas mass of six local group dSphs and the metallicity distribution of Carina dSph galaxy by means of a detailed chemical evolution model that adopts low SFR (efficiencies in the range $\nu = 0.01-0.5 \;Gyr^{-1}$, with the exception of Sagittarius, which is characterised by $\nu = 1.0-5.0\;Gyr^{-1}$) 
and intense galactic winds (with high rates: 4 to 13 times the star formation rate - $w_i$ = 4 - 13). In their scenario, the pattern of the abundance ratios is defined by the low SFR and by the effects of the intense galactic winds that lower the SF further. Moreover, the winds are able to remove a large fraction of the gas reservoir of the galaxy, eventually halting the star formation.

Fenner et al. (2006), on the other hand, argue that the abundance ratios in local dSphs could be reproduced well by a model with low SFR and moderate galactic winds, which would probably not be the only mechanism responsible for removing the gas content of the galaxy. They therefore concluded that an external mechanism should be the main cause of the lack of gas in these galaxies. The removal of gas by external effects, however, also presents several problems. Ram-pressure stripping has been suggested as able to play a crucial role in this removal (van den Bergh 1994). For this to happen, some physical conditions must be attained, such as specific gas densities of the intergalactic medium (IGM) and of the gas in the dwarf galaxy (see for more details Gunn $\&$ Gott 1972). That is not the case for Local Group dSph galaxies, which exhibit 
either higher densities than the ones expected from ram-pressure stripping or are embedded in an IGM that is not dense enough. Besides that, dSphs should always be found in the vicinity of giant galaxies, which is not the case for, at least, Cetus and Tucana dSphs. In that sense, ram-pressure stripping might not be the unique explanation for the depletion of gas in dSphs. Another mechanism commonly invoked is tidal stripping, for those galaxies whose orbits bring them close enough to massive galaxies. In that case it would be very difficult to explain the most
isolated dSphs, again like the extreme cases of Tucana and Cetus.

In this work we show not only that the abundance ratios and gas mass are reproduced by models with galactic winds, but also that the observed SMDs of dSphs can only be obtained if intense winds are taken into account. The very sharp decrease at the high-metallicity tail of the SMDs in local dSph (Bellazzini et al. 2002; Koch et al. 2006) is generally attributed to the cessation of the SF. Only a sudden, although not instantaneous, interruption of the SF would cause a decrease similar to the one observed and the galactic winds are very promising candidates for explaining this feature.

The paper is organised as follows: in Sect. 2 we present
the observational data concerning the SMDs of Draco and Ursa Minor dSphs; in Sect. 3 we summarise the main characteristics of the adopted chemical evolution models, such as the SF and the galactic winds prescriptions; in Sect. 4 the predictions of our models are compared to the photometrically-derived SMDs, and the results are discussed.
Finally in Sect. 5 we draw some conclusions. All elemental
abundances are normalised to the solar values
([X/H] =  log(X/H) - log(X/H)$_{\odot}$)  as measured by 
Grevesse $\&$ Sauval (1998).

\section{Data sample}

We compared the SMDs predicted by our chemical evolution models to the ones derived by Bellazzini et al. (2002). In that work, the stellar metallicities of Draco and Ursa Minor stars were inferred, using two different metallicity scales (Zinn $\&$ West 1984 - ZW - and Carretta $\&$ Gratton 1997 - CG), from the colours of RGB old-population stars. The photometric metallicity distributions were constructed by interpolating in colour between the RGB ridge lines of a grid of metal templates for stars in the magnitude range $-2.9 \le M_I \le -3.9$. The authors suggest that, even though the singe metallicities are not very accurate, the main features (such as position of the peak, the shape of the distribution, and the dispersion in metallicity) of the derived SMDs are quite robust, since the SMD is based on a large number of stars (roughly a hundred). One extra possible source of uncertainty resides in the fact that the metallicity spread derived with this procedure is the convolution of the intrinsic metallicity spread with the colour spread due to observational scatter, which, in turn, depends on the mean metallicity. 

Even though the authors claim that there is good agreement between the photometric metallicity estimates with the high-resolution spectroscopic abundances from Shetrone et al. (2001), it should be mentioned that the metallicity range of their SMDs does not cover the entire range of spectroscopic metallicities (given by [Fe/H]). For both galaxies there are stars with spectroscopically-derived [Fe/H] (from $\sim$ -2.5 to -2.9 dex) much below the lower limit of the photometric SMDs ($\sim$ -2.1 dex for Draco and $\sim$ -2.3 dex for Ursa Minor), mainly due to the insensitive of the method to stars with [Fe/H] $<$ -2.5 dex. Consequently, the comparisons between those SMDs with model predictions should be focused on the general features of the distribution, giving minor importance to minor details.

The overall shape of the photometric SMD of Draco and Ursa Minor galaxies are very similar, but some differences are visible. At first glance, one notices that the SMDs derived with the CG scale for both galaxies are located at metallicities higher than the ones adopting the ZW scale, even though the dispersion in metallicity is almost the same on CG and ZW scales (the spread in SMD with the CG scale is 0.1 dex larger in both galaxies). Differences between the SMDs of the two galaxies concerning the peak of the distributions and their spread in metallicity can also be noticed. Draco seems to be slightly  poorer in metal (the peak is at lower [Fe/H] -$\sim$- 0.15 dex) than Ursa Minor, but with a slightly larger metallicity spread ($\sim$ 0.2 dex larger). The increase in the SMD at the metal-poor tail is smoother for Draco than for Ursa Minor, but the decrease in the metal-rich tail is more intense in the case of Draco.
These small differences suggest a similar scenario for the chemical evolution histories of these two galaxies, but each one with its own particular details. The fact that Draco is poorer in metal than Ursa Minor, but has a large metallicity spread
could, for example, be related to the former galaxy being characterised by a longer SF activity but occurring at a lower efficiency. On the other hand, the general evolutionary scenario for both galaxies should be the same, as indicated by the general features of the observations, as we intend to show in the next sections.
%%%%%%%%%%%%%%%%%%%%%%%%%%%%%%%%%%%%%%%%%%%%%%%%%%%%%%%%%%%%%%%%%%%
\begin{table*}
%\begin{flushleft} 
\begin{center}\scriptsize  
\caption[]{Models for dSph galaxies.}
\begin{tabular}{lcccccccccc}  
\hline\hline\noalign{\smallskip}  
galaxy &$\nu(Gyr^{-1})$ &$w_i$
&n &t($Gyr$) &d($Gyr$) &$t_{GW}(Gyr)$ &$IMF$ &$\eta_{SNII}$ &$\eta_{SNIa}$ 
&Rate$_{GW}$\\    

\noalign{\smallskip}  
\hline
Draco1a &0.005-0.1  &4 &1 &0 &4 &1.25 &Salpeter &0.03 &1.0 &SFR\\
Draco1b &0.005-0.1  &4 &1 &0 &4 &1.97 &Salpeter &0.03 &0.5 &SFR\\
Draco2 &0.005-0.1  &4 &1 &0 &4 &1.25/1.97 &Salpeter &0.03 &1.0/0.5 &SNe Rate\\

Draco3 &0.005-0.1  &4 &1 &0 &4 &-- &Salpeter &0.03 &1.0 &No Wind\\
Ursa Minor1 &0.05-0.5 &10 &1 &0 &3 &0.43 &Salpeter &0.03 &1.0 &SFR\\
Ursa Minor2 &0.05-0.5 &10 &1 &0 &3 &0.43 &Salpeter &0.03 &1.0 &SNe Rate\\
Ursa Minor3 &0.05-0.5 &10 &1 &0 &3 &-- &Salpeter &0.03 &1.0 &No Wind\\
\hline\hline
\end{tabular}
\end{center}
%\end{flushleft}
\end{table*} 
%%%%%%%%%%%%%%%%%%%%%%%%%%%%%%%%%%%%%%%%%%%%%%%%%%%%%%%%%%%%%%%%

\section{Models} 

We compared the observed photometric SMDs of Draco and Ursa Minor to the distributions predicted by the models for these two galaxies using the same models as described in LM03 and LM04. 
These models have already reproduced several [$\alpha$/Fe] ratios, the 
[Ba/Fe] and [Eu/Fe] ratios very well (Lanfranchi, Matteucci $\&$
Cescutti 2006b - LMC06b), as well as the current total mass and gas mass observed in these two galaxies.

According to these models, the dSphs form 
through a continuous and fast infall of pristine gas until 
a mass of $\sim 5 \cdot 10^8 M_{\odot}$ is accumulated. 
The SFH of each galaxy is given by the analysis of observed CMDs that suggest a unique long ($t \sim 3-4 \;Gyr$) episode of SF in both systems (Hernandez et al. 2000; Aparicio et al. 2001; Carrera et al. 2002; Dolphin et al. 2005). The duration of the episodes taken from the CMDs should, however, be viewed in the models as the maximum possible duration. The time interval in which the SF is active is imposed as an initial value in the models. If there was no galactic wind, that would be the actual duration of the SF, but since there is an intense wind, the gas is removed quickly. As a consequence, the SFR also decreases quickly reaching, in some cases, negligible values before the epoch imposed as an initial condition for the SF cessation. In the case of Draco and Ursa Minor, the SF in the models began around 14 Gyr ago and lasted for 3 Gyr (Ursa Minor) and 4 Gyr (Draco), as suggested by observations, but the epoch of the maximum activity is determined by the occurrence of the galactic wind: normally the time just before the wind starts ($\sim$ 1.5 - 2 Gyr after the onset of the SF).

The galactic wind is actually a vital ingredient in the model by playing a crucial role in the evolution of these galaxies. When the thermal energy of the gas is equal to or larger than its binding energy, a galactic wind 
is initiated (see for example Matteucci
$\&$ Tornamb\'e 1987). The thermal energy strongly depends on the assumptions regarding the thermalization efficiency of supernovae of both types and of stellar winds (Bradamante et al. 1998). In this work we assume the thermalization efficiencies suggested by Recchi et al. (2001) as standard values, i.e. $\eta_{SNeIa} = 1.0$, $\eta_{SNeII} = 0.03$, $\eta_{SW} = 0.03$ for SNeIa, SNeII, and stellar winds, respectively. The binding energy, on the other hand, is strongly influenced by assumptions concerning the presence and distribution of dark matter (Matteucci 1992). A diffuse ($R_e/R_d$=0.1, where $R_e$ is the effective radius of the galaxy and $R_d$ is 
the radius of the dark matter core) but massive 
($M_{dark}/M_{Lum}=10$) dark halo has been assumed for each galaxy. 
This particular configuration allows the development of a galactic wind in these small systems without destroying them. 

The galactic winds are intense (with rates 4-13 times the SFR) in the models for the dSphs, in order to reproduce the fast decrease observed in the [$\alpha$/Fe] ratios as functions of [Fe/H], and also to remove a large fraction of the gas content of the galaxy. A possible justification for intense galactic winds in dSphs comes from the fact that their potential well is not as deep as in the case of the dwarf irregular galaxies and they are more extended (Mateo 1998; Guzm\'an et al 1998;
Graham \& Guzm\'an 2003). 

The model allows us to follow the evolution of the 
abundances of several chemical elements in detail, starting from 
the matter reprocessed by the stars and restored into the 
interstellar medium (ISM) by stellar winds and type II and Ia supernova
explosions.

The main assumptions of the model are:

\begin{itemize}

\item
one zone with instantaneous and complete mixing of gas inside
this zone;

\item
no instantaneous recycling approximation, i.e. the stellar 
lifetimes are taken into account;

\item
the evolution of several chemical elements (H, D, He, C, N, O, 
Mg, Si, S, Ca, Fe, Ba, and Eu) followed in detail;

\item
the nucleosynthesis prescriptions including the yields of 
Nomoto et al. (1997) for type Ia supernovae, Woosley $\&$
Weaver (1995) (with the corrections suggested by 
Fran\c cois et al., 2004) for massive stars ($M > 10 M_{\odot}$), 
van den Hoek $\&$ Groenewegen (1997) for intermediate mass stars 
(IMS), and for Ba and Eu the ones described in Cescutti et al. (2006) and 
Busso et al. (2001). 
\end{itemize}

We adopt the basic equations of chemical evolution as described 
in Tinsley (1980) and Matteucci (1996). The prescriptions for the SF (which follows a Schmidt law - Schmidt 1963), initial mass function (IMF - Salpeter 1955), infall, and galactic winds are the same as in LM03 and LM04. The type Ia SN progenitors are assumed to be white dwarfs in binary systems according to the formalism originally developed by Greggio $\&$ Renzini (1983) and Matteucci $\&$ Greggio (1986).
The main parameters adopted for each galaxy, together with the predicted time for the occurrence of a galactic wind, $t_{GW}$, can be seen in Table 1, where $\nu$ is the star-formation 
efficiency, $w_i$ is the wind efficiency,  $n$, $t$ and $d$ 
are the number, time of occurrence and duration of the SF 
episodes, respectively, $t_{GW}$ the time of the occurrence of the 
galactic wind, $\eta_{SNII}$ and $\eta_{SNIa}$ are the thermalization efficiencies for SNe II and SNe Ia, and Rate$_{GW}$ specifies the proportionality of the galactic wind rate. In particular, the efficiencies of the SF are low ($\nu=0.005-0.1\;Gyr^{-1}$ for Draco and $\nu=0.05-0.5\;Gyr^{-1}$ for Ursa Minor), whereas the galactic wind efficiencies are high.

In order to verify the effects of the galactic winds on the evolution of the 
dSphs, we tested three different prescriptions for the winds. The first one 
(model 1) is exactly the same as in previous papers (LM03, LM04). 
The rate of gas loss, $\dot G_{iw}$, is assumed to be proportional to 
the SFR, $\psi(t)$, through a proportionality constant, $w_i$, which is a free
parameter describing the efficiency of the galactic wind for any given element
$i$:

\begin{eqnarray}
 \dot G_{iw}\,=\,w_{i} \, \psi(t).
\end{eqnarray}

\noindent
In this scenario, the thermalization efficiencies for SN explosions of both types (SNe Ia, SNe II) and stellar winds are the ones suggested by Recchi et al. (2001).

The second model (model 2) adopts the scenario for galactic winds suggested by Romano et al. (2006), in which the rate of the gas loss is proportional to the rate of SN explosions and to the total amount of gas at that present time, instead of being proportional to the SFR.  For each element $i$ we can write

\begin{eqnarray}
  \dot G_{iw}\,=\,
    \beta_i ([\eta_{\mathrm{SNII}} R_{\mathrm{SNII}}(t) + 
              \eta_{\mathrm{SNIa}} R_{\mathrm{SNIa}}(t)]) X_{i}(t) {\mathcal{G}}(t),
\end{eqnarray}

\noindent
where $\beta_i$ is the equivalent of $w_i$ in model 1, $\dot G_{iw}$ is the rate of gas loss, $R_{\mathrm{SNII}}$ and $R_{\mathrm{SNIa}}$ are the rates of SNe II and SNIa, and $X_{i}(t) {\mathcal{G}}(t)$ is the gas fraction of element $i$ . The efficiencies of thermalization suggested by Romano et al. (2006) are different: $\eta_{SNeIa} = 1.0$, $\eta_{SNeII} = 0.20$, $\eta_{SW} = 0.03$. This difference leads to a galactic wind developing sooner than in the previous case, due to the higher efficiency of thermalization for 
SNe II. It should be mentioned that this scenario was suggested in order to reproduce the properties of a dwarf irregular galaxy, which is normally characterised by a large amount of neutral gas, contrary to the dSphs studied here. In order to reproduce the total absence of gas in dSph and the observed abundance ratios, we changed the efficiency of the wind, making it higher.

In the last scenario (model 3), we adopt a model with no galactic wind. Since there is no wind, the gas content of the galaxy should be removed by external factors, such as ram-pressure or tidal stripping, at the end of the SF. In this model, the duration of the SF episode is the maximum one, as given by the observed CMDs and imposed as initial value in the models.

In these three models only the prescriptions of the wind and the thermalization efficiencies differ from one model to another. All the other parameters are kept the same (see Table 1).

\section{Results}

We compared the predictions of the three models, each with a specific formulation for the galactic wind, for Draco and Ursa Minor with the SMDs derived by photometric observations. Even though the uncertainties in the derived SMDs affect some small details of the distribution, the overall shape is robust and so are the general aspects of the comparison. On the other hand, major differences could help in constraining the models of chemical evolution regarding both the SFH and the prescriptions of the galactic winds as we show in the next sections.

\subsection{Draco dSph galaxy} 

The dSph galaxy Draco is characterised by an ancient stellar population that would have been almost entirely formed from 10 to 14 Gyr ago (Aparicio et al. 2001; Dolphin et al. 2005). The 
mean stellar abundance is low ([Fe/H] $\sim$ -1.7 dex) and the [$\alpha$/Fe] ratios exhibit a sharp decrease at [Fe/H] $\sim$ -1.6 dex (Shetrone et al. 2003). This sharp decrease was explained by LM03 and LM04 as a consequence of the effect of an intense galactic wind on the already low SFR of the galaxy ($\nu$ = 0.005 - 0.1 $Gyr^{-1}$, $w_i$ = 4). When the wind develops, it removes a large fraction of the gas reservoir that fuels the SF thus decreasing the SFR. With an almost negligible SFR, the injection of oxygen (and $\alpha$ elements in general) into the ISM is almost halted, since the main producers of this element - SNe II - have a short lifetime. Iron, on the other hand, is injected into the ISM in a much larger timescale, since the progenitors of SNe Ia, have lifetimes ranging from 35 Myr to several Gyr. Therefore, after the development of the galactic wind, the [$\alpha$/Fe] ratios suffer an intense decrease. The same scenario also holds for [Ba/Fe] and [Eu/Fe] (LMC06b). 

%%%%%%%%%%%%%%%%%%%%%%%%%%%%%%%%%%%%%%%%%%%%%%%%%%%%%%%%%%%%%%%%%%%
\begin{figure}
\centering
%\vspace{0.001cm}
\includegraphics[height=8cm,width=8cm]{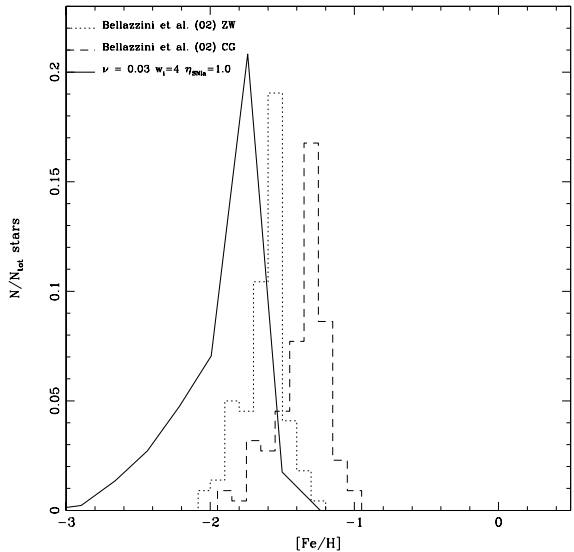}
\caption[]{The observed stellar metallicity distribution of Draco compared to the prediction of model 1a. The dashed line corresponds to the SMD derived with the CG scale, and the dotted line represents the SMD derived with ZW scale. The solid line corresponds to the prediction of model 1a ($\nu = 0.05\;Gyr^{-1}$, $w_i = 4\;Gyr^{-1}$)
.

} 
\end{figure}
%%%%%%%%%%%%%%%%%%%%%%%%%%%%%%%%%%%%%%%%%%%%%%%%%%%%%%%%%%%%%%%%%%%
The SMD of Draco is also affected by the low SF efficiency and by the high wind efficiency and its influence on the SFR, as we shall see in the comparisons with the observed data. In Fig. 1, we compare the observed SMD from Bellazzini et al. (2002) with the predictions of the model 1a. As one can see, when adopting the same model, with the same parameters as in LM04, the predicted SMD fails to reproduce the observed one, even the one with ZW scale, which exhibits lower metallicities. The predicted SMD is located at low metallicities with a peak at lower [Fe/H] (predicted peak at $\sim$ -1.8 dex, observed at $\sim$ -1.6 dex). The lower value in the predicted SMD is a consequence of the influence of the galactic wind on the SF. The galactic wind, in this model, starts at a metallicity of [Fe/H] $\sim$ -1.6 dex, so the number of stars formed with metallicities above that is small. This fact not only causes an early decrease (when compared to observations) in the high-metallicity tail of the distribution, but also produces a peak at lower metallicities. If the wind, on the other hand, develops later, at higher metallicities, then the high-metallicity tail would be shifted to higher values, as it would be the peak of the distribution. 

The epoch of the occurrence of the galactic wind depends mainly on the SFR and on the thermalization efficiencies adopted for SNe II and SNe Ia. The SF efficiency is adopted in such a way that the predicted abundance ratios reproduce the observed values, so it cannot be changed considerably without losing the agreement obtained in the previous works (LM03, LM04). The thermalization efficiencies adopted, on the other hand, are based on hydrodynamical simulations, in particular the results of Recchi et al. (2001). These authors studied the hydrodynamical and chemical evolution of dwarf irregular galaxies and concluded that, whereas the thermalization efficiency for SNe II is low (only 3 $\%$), it could be very high (100 $\%$) for SNe Ia. This is explained by the fact that, when SNe II explode, the ISM is still cold and dense, and the major fraction of the energy is lost by cooling.
Later, when the SNe Ia start exploding, the medium is hotter and less dense than before, so the energy released is totally converted  into thermal energy of the gas. This scenario reproduces the environment of starburst galaxies suffering several short (few hundred Myr) and intense bursts of SF very well, but we are interested here in galaxies that are normally characterised by long (several Gyr) episodes of SF with lower efficiencies. In such a scenario it could be possible that SNe Ia thermalization efficiency is not as high, since it depends on physical conditions such as density of the medium, the mass of the SF region, and others (Bradamante et al. 1998; Melioli $\&$ Dal Pino 2004), which are different in dSph galaxies. In that sense, we also ran models with different $\eta_{SNIa}$ in order to test its effects on the predicted SMD.

%%%%%%%%%%%%%%%%%%%%%%%%%%%%%%%%%%%%%%%%%%%%%%%%%%%%%%%%%%%%%%%%%%%
\begin{figure}
\centering
%\vspace{0.001cm}
\includegraphics[height=8cm,width=8cm]{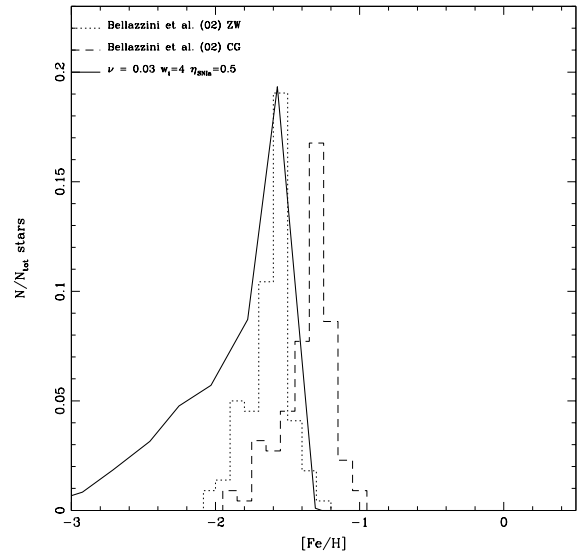}
\caption[]{The observed stellar metallicity distribution of Draco compared to the prediction of model 1 with $\eta_{SNIa}=0.5$ (model 1b). The dashed line corresponds to the SMD derived with the CG scale, and the dotted line represents the SMD derived with ZW scale. The solid line corresponds to the prediction of model 1 ($\nu = 0.05\;Gyr^{-1}$, $w_i = 4\;Gyr^{-1}$)
.

} 
\end{figure}
%%%%%%%%%%%%%%%%%%%%%%%%%%%%%%%%%%%%%%%%%%%%%%%%%%%%%%%%%%%%%%%%%%%

In Fig. 2 , the SMD predicted by the model 1 is shown, but with $\eta_{SNIa} = 0.50$ (model 1b), compared to the observed values. When a lower $\eta_{SNIa}$ is adopted, the wind is developed later in time and, consequently, at higher metallicities, shifting the SMD toward higher [Fe/H]. This scenario provides an SMD that is in better agreement with the observed SMD derived with the ZW scale. The peaks of the predicted and observed metallicities are now at the same position ([Fe/H] $\sim$ -1.6 dex) and the fast decrease of the high-metallicity tail is similar in both distributions. As mentioned before, this decrease is a consequence of the effect of the intense wind on the SFR, which decreases to negligible values in a short timescale preventing more stars with higher metallicities being formed. The only difference between theory and observations is found in the low-metallicity tail, below [Fe/H] $\sim$ -2.0 dex. The model appears to overpredict the number of metal-poor stars, thus leading to a problem similar to the so-called "G-dwarf problem" in the Milky Way. In the Milky Way, the problem was solved simply by adding infall to the models. Here, on the other hand, the reason for the discrepancy between model prediction and observations could be related to the procedure adopted to derive the observed SMD. As mentioned in sect. 2, the observed SMD is constructed by deriving [Fe/H] from the colours of the stars, a technique that leads to some uncertainties especially regarding the spread of the distribution. Besides that, the procedure is insensitive to stars with [Fe/H] $<$ -2.5 dex, which are actually present in spectroscopic studies (Shetrone et al. 2003). In that sense, the absence of stars that are more metal-poor than [Fe/H] $\sim$ -2.0 dex in the observed SMD does not reflect the real scenario, since they do exist in Draco dSph galaxy. The number of these stars is not known, however, so it is not possible yet to argue that the model predictions are correct. Only spectroscopic SMD would help to impose tighter constraints. In conclusion, the general features (e.g. position of the peak, decrease at the high-metallicity tail, shape) of the predicted SMD are in agreement with the observations, suggesting that the general scenario adopted for the evolution of Draco dSph galaxy is reasonable, especially concerning the main parameters such as the intensity of the galactic wind and the SF efficiency.

In order to analyse different scenarios for the galactic winds, model 2 was first adopted, in which the rate of the galactic wind is proportional to the rate of SNe of both types, instead of the SF rate. 
In Fig. 3, the predictions from model 2 are compared to the observed SMD. We tested model 2 with two different thermalization efficiencies for SNe Ia (model 2a: $\eta_{SNIa}$ = 1 and model 2b: $\eta_{SNIa}$ = 0.5). The results for this model are almost the same as for model 1, since the distributions predicted by both models are very similar. Again, the model with high SNe Ia thermalization efficiency fails to reproduce the data, by predicting an SMD that is too skewed toward lower values of [Fe/H] when compared to the observations. When $\eta_{SNIa}$ is reduced to 50 $\%$, the wind is delayed and the agreement of the predicted SMD with the observed ZW distribution is quite good. In particular, the general features of the distribution (shape, position of the peak, decrease at the high-metallicity tail) fit the observed ones, whereas there is a discrepancy at low [Fe/H]. The total absence of stars more metal-poor than [Fe/H] = -2.0 dex in the observed SMD is not real, given the fact that there are stars observed in this metallicity range (Shetrone et al. 2003). Therefore, we do not take the discrepancy at low [Fe/H] into account. The major difference between models 1 and 2 is the remaining mass of gas in the galaxy. When adopting model 1, the predicted present-day gas mass of the galaxy is below the lower limits inferred by observations ($<$ $10^4\;M_{\odot}$), due to the intense galactic winds that swept out a large fraction of the gas . In model 2, on the other hand, the wind is not as intense as in model 1, since the rate of gas loss is proportional to the rate of SNe of both types. A less intense wind removes a lower fraction of the gas, leaving a large amount of gas at the present time. The final gas mass in model 2 is much higher (order of 100) than inferred by observations (Table 2), therefore requiring an external mechanism to remove the remaining gas of the galaxy in order to fit observations.

%%%%%%%%%%%%%%%%%%%%%%%%%%%%%%%%%%%%%%%%%%%%%%%%%%%%%%%%%%%%%%%%%%%
\begin{figure}
\centering
%\vspace{0.001cm}
\includegraphics[height=8cm,width=8cm]{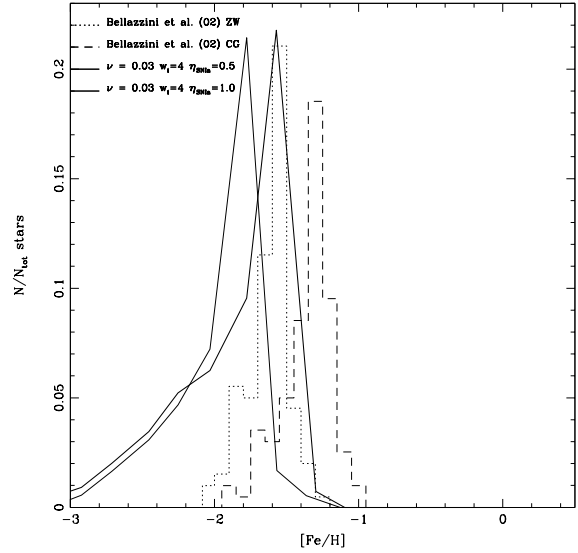}
\caption[]{The observed stellar metallicity distribution of Draco compared to the prediction of model 2. The dashed line corresponds to the SMD derived with the CG scale, and the dotted line represents the SMD derived with ZW scale. The solid line corresponds to the prediction of model 2 ($\nu = 0.05\;Gyr^{-1}$)
.

} 
\end{figure}
%%%%%%%%%%%%%%%%%%%%%%%%%%%%%%%%%%%%%%%%%%%%%%%%%%%%%%%%%%%%%%%%%%%

The comparison between the observed SMDs and the predictions of model 3, in which there is no occurrence of galactic winds, is shown in Fig. 4. In a no wind scenario, the SFR is higher than in the wind cases. In this case, at the end of the SF phase, an external mechanism, such as ram-pressure stripping, would produce a sudden total removal of gas. In a galactic wind case, the decrease in the SFR is also intense but less abrupt. In fact, the galactic wind removes the gas content of the galaxy more slowly. Initially, the fraction of gas content swept out of the galaxy by the wind is high, but this fraction decreases in time (see Fig. 5 in LM03). Consequently, the SFR is substantially reduced soon after the wind is developed, but this decrease becomes softer later on as the amount of gas lost by winds is diminished.
The time at which the SF is stopped in the case of a model without wind is given by the SFHs inferred by the analysis of the observed CMD (at 4 Gyr - Dolphin et al. 2005; Aparicio et al. 2001). Since the SF in such a model is higher than in the wind model, the number of stars formed continues to increase with increasing metallicity, leading to a higher number of stars with higher [Fe/H] and to a predicted SMD at higher [Fe/H] than the SMD derived with both metallicity scales (CG and ZW). The decrease at the high-metallicity tail of the predicted SMD is also much steeper than the observed one due to the instantaneous suppression of the SF.
As a consequence, the predicted SMD exhibits a peak at higher [Fe/H] and with a very intense decrease of the high-metallicity tail, at variance with the observed values.

Therefore, it is clear that, in order to reproduce the observed SMD of Draco, an intense continuous galactic wind is required, no matter whether its rate is proportional to the SFR (model 1) or to the SN rate (model 2), even though model 1 reproduces the observationally estimated final masses (gas and total) better.

%%%%%%%%%%%%%%%%%%%%%%%%%%%%%%%%%%%%%%%%%%%%%%%%%%%%%%%%%%%%%%%%%%%
\begin{figure}
\centering
%\vspace{0.001cm}
\includegraphics[height=8cm,width=8cm]{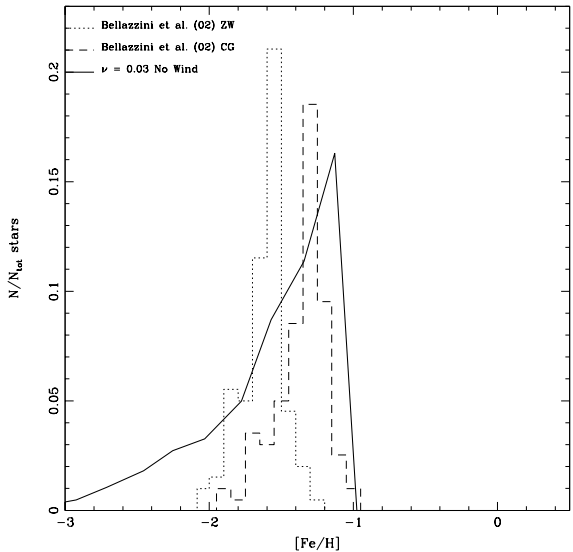}
\caption[]{The observed stellar metallicity distribution of Draco compared to the prediction of model 3, with no galactic winds. The dashed line corresponds to the SMD derived with the CG scale, and the dotted line represents the SMD derived with ZW scale. The solid line corresponds to the prediction of model 3 ($\nu = 0.03\;Gyr^{-1}$)

} 
\end{figure}
%%%%%%%%%%%%%%%%%%%%%%%%%%%%%%%%%%%%%%%%%%%%%%%%%%%%%%%%%%%%%%%%%%%

%%%%%%%%%%%%%%%%%%%%%%%%%%%%%%%%%%%%%%%%%%%%%%%%%%%%%%%%%%%%%%%%%%%
\begin{table}
%\begin{flushleft} 
\begin{center}\scriptsize  
\caption[]{The present-day $(M_{HI}/M_{Tot})$ predicted by the models for Draco and Ursa Minor dSph galaxies compared to the observed values given in Mateo (98).}
\begin{tabular}{lcc}  
\hline\hline\noalign{\smallskip}  
galaxy &$(M_{HI}/M_{Tot})^{obs}$ &$(M_{HI}/M_{Tot})^{pred}$\\    

\noalign{\smallskip}  
\hline
Draco1a &$<$0.001  &0.0013\\
Draco1b &$<$0.001  &0.00038\\
Draco2 &$<$0.001  &0.205/0.239\\
Draco3 &$<$0.001  &0.92\\
Ursa Minor1 &$<$0.002 &0.0003\\
Ursa Minor2 &$<$0.002 &0.020\\
Ursa Minor3 &$<$0.002 &0.72\\
\hline\hline
\end{tabular}
\end{center}
%\end{flushleft}
\end{table} 
%%%%%%%%%%%%%%%%%%%%%%%%%%%%%%%%%%%%%%%%%%%%%%%%%%%%%%%%%%%%%%%%

\subsection{Ursa Minor}

Ursa Minor, as for Draco, is believed to be characterised by an ancient stellar population that would have been entirely formed between 11 and 14 Gyr ago. The SF would have occurred in one long episode ($\sim$ 3 Gyr) of activity (Dolphin et al. 2005; Carrera et al. 2002) with a low rate ($\nu$ = 0.05 - 0.5 $Gyr^{-1}$ - LM04).

%%%%%%%%%%%%%%%%%%%%%%%%%%%%%%%%%%%%%%%%%%%%%%%%%%%%%%%%%%%%%%%%%%%
\begin{figure}
\centering
%\vspace{0.001cm}
\includegraphics[height=8cm,width=8cm]{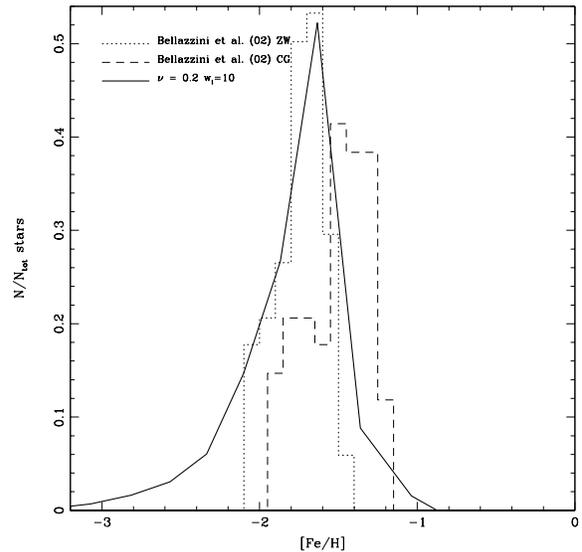}
\caption[]{The observed stellar metallicity distribution of Ursa Minor compared to the prediction of model 1. The dashed line corresponds to the SMD derived with the CG scale, and the dotted line represents the SMD derived with ZW scale. The solid line corresponds to the prediction of model 1 ($\nu = 0.1\;Gyr^{-1}$, $w_i$ = 10)
.

} 
\end{figure}
%%%%%%%%%%%%%%%%%%%%%%%%%%%%%%%%%%%%%%%%%%%%%%%%%%%%%%%%%%%%%%%%%%%

The comparison between the observed SMD and the one predicted by the same chemical evolution model for Ursa Minor as in LM04 and LMC06b is shown in Fig. 5. With no modifications in the main parameters of model 1 for Ursa Minor (see Table 1), the predicted SMD fits the ZW SMD very well: the shape is very similar to the observed one, the location of the peak is almost the same ([Fe/H] $\sim$ -1.6 - -1.7 dex), and the decrease at the high-metallicity tail is intense in both cases. One can notice that Ursa Minor's SMD (both observed and predicted) is more symmetric (in the low- and high-metallicity tails) when compared to Draco's, in the sense that the decrease in the high-metallicity tail is not as intense as for Draco. This is due to the fact that the SF efficiency for Ursa Minor best model is much higher than for Draco's best model. With a higher SFR, the number of stars formed after the onset of the wind decreases less  than when the SF efficiency is lower, giving rise to a higher number of stars formed with metallicities higher than the one reached when the wind develops. As in the case of Draco, one can notice that the number of metal-poor stars predicted by the model is higher than the observed one. Once more, this fact is related to the method used to derive the observed SMD, which is insensitive to metallicities lower than [Fe/H] = -2.5 dex. Since there are stars in this metallicity range also observed in Ursa Minor (Shetrone et al. 2003; Sadakane et al. 2004), this does not seem to be a problem. Further accurate observations at lower metallicities would impose constraints on the relative number of metal-poor stars present in these galaxies.

%%%%%%%%%%%%%%%%%%%%%%%%%%%%%%%%%%%%%%%%%%%%%%%%%%%%%%%%%%%%%%%%%%%
\begin{figure}
\centering
%\vspace{0.001cm}
\includegraphics[height=8cm,width=8cm]{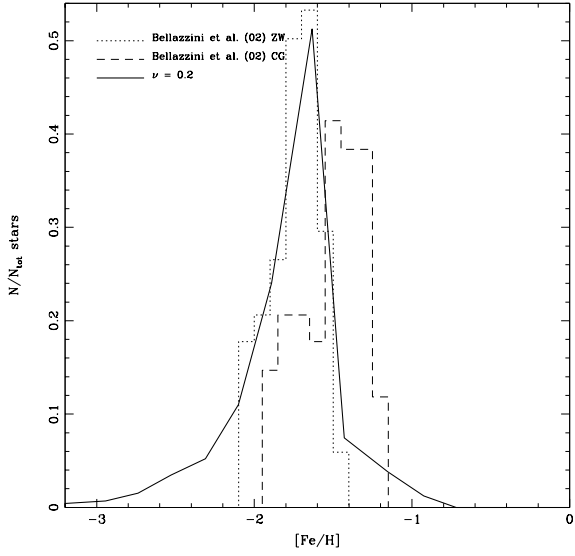}
\caption[]{The observed stellar metallicity distribution of Ursa Minor compared to the prediction of model 2. The dashed line corresponds to the SMD derived with the CG scale, and the dotted line represents the SMD derived with ZW scale. The solid line corresponds to the prediction of model 2 ($\nu = 0.1\;Gyr^{-1}$)
.

} 
\end{figure}
%%%%%%%%%%%%%%%%%%%%%%%%%%%%%%%%%%%%%%%%%%%%%%%%%%%%%%%%%%%%%%%%%%%

In the case of Ursa Minor, contrary to Draco's case, the maximum efficiency of thermalization for SNe Ia ($\eta_{SNIa}=1.0$) adopted in the model provides an SMD in very good agreement with the observed one. This fact suggests that, even though very similar, the dSph galaxies might exhibit differences, like the ones seen in the SFH (Dolphin et al. 2005), in the abundance ratios (Venn et al. 2004) and in the SF and galactic wind efficiencies (LM04). Our model suggests that the differences between Draco and Ursa Minor are in the SF and galactic wind efficiencies and in the efficiency of thermalization for SNe Ia. This could be related to different physical properties of their ISM in the early stages of their evolution. 

In order to analyse the effects of different prescriptions for the galactic winds on the SMD of Ursa Minor, as was done in Draco's case, we also compared Bellazzini's SMD with the predictions of model 2 for Ursa Minor, which adopts a wind rate proportional to the SN rate. In Fig. 6 one can see that the comparison is very similar to the one adopting model 1. With a galactic wind rate proportional to the SN rate (instead of SFR), it is also possible to reproduce the SMD in the ZW metallicity scale very well: the peak is at the same position ([Fe/H] $\sim$ -1.6 - -1.7 dex), the shapes are very similar, the decrease at the high-metallicity tail occurs at the same point and with very similar slopes.

The only small difference arising from the comparison of models 1 and 2 is that the decrease in the high-metallicity tail seems to be more intense at the beginning and then softer at the highest metallicities in model 2. This difference, however, is not significant enough to distinguish between the models, given the fact that the small details of the observed distribution might not be accurate enough. On the other hand, the predictions of model 2 for the final gas mass do not exhibit an agreement with observations as good as model 1 predictions, similar to what has happened with Draco. Consequently, these results suggest that the best scenario for the wind in Ursa Minor (and also Draco) is the one adopted in model 1 (and in all previous models), i.e. a very intense galactic wind with a rate proportional to the SFR.

%%%%%%%%%%%%%%%%%%%%%%%%%%%%%%%%%%%%%%%%%%%%%%%%%%%%%%%%%%%%%%%%%%%
\begin{figure}
\centering
%\vspace{0.001cm}
\includegraphics[height=8cm,width=8cm]{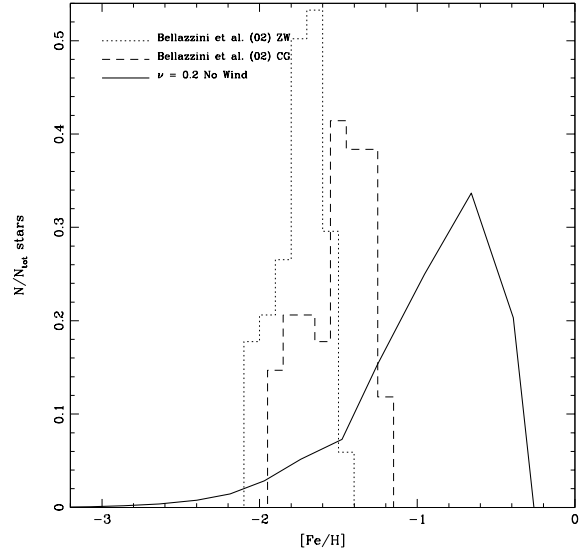}
\caption[]{The observed stellar metallicity distribution of Ursa Minor compared to the prediction of model 3, with no galactic winds. The dashed line corresponds to the SMD derived with the CG scale, and the dotted line represents the SMD derived with ZW scale. The solid line corresponds to the prediction of model 3 ($\nu = 0.1\;Gyr^{-1}$)
.

} 
\end{figure}
%%%%%%%%%%%%%%%%%%%%%%%%%%%%%%%%%%%%%%%%%%%%%%%%%%%%%%%%%%%%%%%%%%%

A model with no galactic winds (thick solid line in Fig. 7) fails to reproduce the observed SMDs also in the case of Ursa Minor. One can notice in Fig. 7 that the SMD predicted by model 3, with no galactic wind, is located at much higher metallicities than the ones observed.

Since the efficiency of the SF is higher in Ursa Minor than in Draco, the number of stars formed with high metallicities is even larger, producing a peak in the SMD at much higher [Fe/H]. In such an extreme case, almost the whole predicted SMD is dislocated to higher [Fe/H] with respect to the observed values, causing a total disagreement between observed data and models. A model with no galactic wind is not only completely unable to reproduce the observed SMD, but it is
also inadequate for explaining the [$\alpha$/Fe] and [Eu/Fe] abundance ratios, especially those with the lowest values at [Fe/H] $<$ $\sim$ -1.6 dex (see the case of Sagittarius in LMC06a). These facts strongly suggest that an intense continuous galactic wind is required to explain several observational features of dSphs, regardless if the wind is proportional to the SFR or SN rate, even though the results of our models indicate that the first case is favoured, since the gas is totally removed by the wind without any need to invoke an additional mechanism.

\section{Discussion}
 
The results of our chemical evolution models for dSph galaxies adopting different prescriptions for the galactic winds, compared to photometric stellar metallicities distributions, suggest that the observed data can only be reproduced if intense galactic winds are taken into account in the evolution of these galaxies. It would be interesting, now, to compare these results to other works with different approaches.

Carigi, Hernandez $\&$ Gilmore (2002), Ikuta $\&$ Arimoto (2002), and Fenner et al. (2006), among others, also studied the chemical evolution of dSphs and propose scenarios similar to ours but with some specific differences.
In particular, Carigi et al. (2002), using reliable non-parametric star formation histories as a constraint for four local dSphs, were able to study some features of these galaxies but overpredicted the present-day metallicities. They suggest that this discrepancy could arise from not treating the removal of fresh metals by galactic winds. In our scenario, we are able to reproduce all observed abundance ratios very well, by assuming that a large fraction of the fresh metals produced by SNe are expelled from the galaxy by galactic winds. 
Ikuta $\&$ Arimoto (2002), on the other hand, adopt a closed model, with no galactic winds, and suggest very low SFR (similar to our values) in order to reproduce the abundance ratios observed. They had, however, to invoke external mechanisms to explain the cessation of the SF and the removal of the gas from the galaxies. It should be remembered that in both the above-mentioned works 
the comparison was made with only few data relative to the [$\alpha$/Fe] ratios, and this did not allow them to draw firm conclusions.

Fenner et al. (2006) used a more detailed chemical evolution model with galactic winds and infall to reproduce several abundance ratios considering an evolutionary scenario very similar to ours in which the SFR is very low. The difference is that their galactic wind is not as intense as ours, and they did not try to reproduce the stellar metallicity distribution of the dSph galaxies and compared their predictions to only one galaxy (Ursa Minor). We reproduced several abundance ratios in six local dSph and the SMDs of three galaxies (two in this paper and Carina's in LMC06a).

From the chemical evolution point of view, consequently, a low SFR, coupled with an intense galactic wind, that is the main mechanism responsible for the cessation of the SF, are required to reproduce  the available observed data satisfactorily, especially the abundance ratios and the SMDs. Can this scenario, however, explain other properties of the dSph galaxies? Richer, McCall $\&$ Stasi\'nska (1998) and Tamura, Hirashita  $\&$ Takeuchi (2001) studied the chemical evolution of dSph galaxies by analysing mass-metallicity and mass-luminosity relations, as well as the relation between [O/H] and mean velocity dispersions. Richer et al. 
(1998), using the observed relation between [O/H] and mean velocity dispersions in several types of dynamically hot galaxies, suggest that there is a correlation between these two quantities that can be explained naturally if the chemical evolution proceeded until the energy input from SNe gave rise to a galactic wind, a scenario which is very similar to the one adopted in our models.
They also suggest, however, that galactic winds in dSphs are triggered only by SNe II and that SNe Ia had little impact on the evolution of these systems. This conclusion seems contradicted by the observed low [$\alpha$/Fe] ratios at low [Fe/H] in most dSphs. Their conclusion was based on the [O/Fe] measured in only two PNe in Sagittarius  and Fornax.
A scenario very similar to our own was presented by Tamura, Hirashita \& Takeuchi (2001), where they explain the mass-metallicity relation as due to the SF proceeding at a very low rate until a galactic wind develops and expels the gas out of the system.

One last point to consider is whether the occurrence of intense galactic winds in dSph galaxies is consistent with hydrodynamical simulations. McLow $\&$ Ferrara (1999), Fragile et al. (2003), and Burkert $\&$ Ruiz-Lapuente (1997), among others, examined the possibility of repeated SNe explosions in the interstellar medium of dwarf galaxies to give rise to a galactic wind. 
Burkert $\&$ Ruiz-Lapuente (1997) analysed the heating of the ISM of dSph galaxies by SN explosions and claim that one vital parameter in defining whether a galactic wind would develop is the amount of dark matter considered in the simulations. By adopting a mass of dark matter higher than $10^{10}M_{\odot}$ and a low efficiency for the heating of the ISM by SNe Ia ($\eta_{SNIa}$ = 0.25), they conclude that the energy released by SNe Ia would never be high enough to produce a wind.
This conclusion should be viewed, though, with caution since the value adopted for the efficiency of heat transfer is indeed very low. More recent hydrodynamical simulations suggest a much higher value for $\eta_{SNIa}$, even as high as one (Recchi et al. 2001; Melioli $\&$ Dal Pino 2004), as adopted in our work.

McLow $\&$ Ferrara (1999) argue that galactic winds develop in galaxies with masses between $M = 10^6 M_{\odot} - 10^9 M_{\odot}$. They modeled the effects of SN explosions in dwarf galaxies and conclude that the mass ejection efficiency is very low for galaxies with $M > 10^7 M_{\odot}$, but fresh metals accelerated by SN explosions are almost totally lost from the galaxy. In that sense, galactic winds can occur in dSph galaxies, but they involve only metals and not the total gas.
More recently, Fragile et al. (2003) computed hydrodynamical three-dimensional simulations and considered several scenarios for the dSphs with different numbers of SNe (from 1 to 10) and  
different initial conditions for the gas distribution.
The adopted gas mass and dark matter mass in these simulations are very similar to the ones adopted in our models. These authors conclude that in low-mass systems most of the enriched gas is lost from the cores of the galaxies following multiple SN explosions. They also suggest that the presence of an irregular ISM considerably enhances the loss of enriched material from the galaxy as a whole in both low- and high-mass systems.

It seems, therefore, that the scenario adopted in our work is consistent not only with other chemical evolution models, but also with scenarios that can reproduce other dSph properties such as mass-metallicity, mass-luminosity relations, 
as well as  the relation between [O/H] and mean velocity dispersions, and with results obtained from hydrodynamical simulations.

\section{Summary}

We compared the predictions of several chemical evolutions models, adopting different prescriptions for the galactic winds (including a model with no wind), with the photometrically-derived SMDs of Draco and Ursa Minor dSph galaxies. These comparisons allowed us to analyse the effects of the galactic winds on both the SMDs and the evolution of these galaxies. We adopted the chemical evolution models from LM03 and LM04 for Draco and Ursa Minor, which were able to reproduce some observational features of these two galaxies, such as several [$\alpha$/Fe], [Eu/Fe], [Ba/Fe] ratios and  the gas mass. The models adopt up-to-date nucleosynthesis for intermediate-mass stars and SNe of both types and take into account the effects of SNe on the energetics of the systems. The predictions were compared to the photometric SMDs derived by Bellazzini et al. (2002), which are accurate enough for a global comparison taking the general aspects into account such as the overall shape, the position of the peak, and the high-metallicity tail, but leaving aside more detailed information on the distributions (the fraction of metal-poor stars, the precise metallicity range). In the cases of both Draco and Ursa Minor dSphs, the model that best reproduces the data is the one with an intense galactic wind, with a rate proportional to the SFR. The models with a wind rate proportional to the SN rate  also reproduce the observed SMDs, but do not match the currently estimated gas mass, whereas the models with no galactic winds fail to reproduce all the observed data.

The specific conclusions are the following:

\begin{itemize}

\item
the Draco model with the same parameters as in LM04 and LMC06b ($\nu$ = 0.05 Gyr $^{-1}$, $w_i$ = 4, $\eta_{SNIa}$ = 1.0) is not able to reproduce the observed photometric SMD properly, since it predicts metallicities too low for the whole stellar population of the galaxy, giving rise to a distribution shifted toward lower [Fe/H]. This is due to the fact that the galactic wind is developed at an [Fe/H] too low ($\sim$ -1.8 dex), and after its occurrence the number of stars formed is very low;

\item
a model for Draco with the same parameters as in LM04 and LMC06b, but with lower SNe Ia thermalization efficiency ($\eta_{SNIa}$ = 0.5) produces a galactic wind later in time and at higher metallicities, giving rise to an SMD in agreement with the observations. The overall shape of the distribution, the position of the peak (at [Fe/H] $\sim$ -1.6 dex), and the decrease in the high-metallicity tail match the observed distribution derived with the ZW metallicity scale.

\item
In the case of Ursa Minor, the very same model as the one adopted in LM04 and LMC06b, with the same parameters ($\nu$ = 0.1 Gyr $^{-1}$, $w_i$ = 10, $\eta_{SNIa}$ = 1.0), produces an SMD in agreement with the observed one. The overall shape of the distribution, the position of the peak (at [Fe/H] $\sim$ -1.6 - -1.7 dex), and the decrease in the high-metallicity tail match the observed distribution derived with the ZW metallicity scale. 

\item
There is an apparent discrepancy in both galaxies between observations and predictions in the low-metallicity tail of the SMDs, similar to the G-dwarf problem in the Milky Way. This discrepancy is not taken into account here since the method used to derive the SMD is insensitive to stars with [Fe/H] lower than -2.5 dex, and such stars are present in spectroscopic studies. The real number of these stars is not known yet, so it is impossible to draw firm conclusions on whether or not the model overestimates their number;

\item 
the models with a wind rate proportional to the SN rate predict very similar SMDs to the ones predicted by the models with the wind rate proportional to the SFR and are able to reproduce the observed SMDs in both galaxies. Such models, however, do not reproduce the present-day gas mass of the two galaxies satisfactorily.

\item 
When models with no galactic winds are adopted, the predicted SMDs are located at high metallicities, far away from the observed distributions in both galaxies. The difference in metallicity is more pronounced in the case of Ursa Minor due to higher SF efficiency. The absence of wind leads the SFR to be higher, thus
giving rise to a higher number of metal-rich stars, contrary to what is observed. As a consequence, such models fail to explain not only the observed SMDs, but also the abundance ratios (see the case of Sagittarius in LMC06a) and the present gas mass, strongly suggesting the necessity of galactic winds in the evolution of the local dSphs.

\end{itemize}

\section*{Acknowledgments}
G.A.L. thanks the Dipartimento di 
Astronomia-Universit\'a di Trieste for its hospitality and full support during his stay.
G.A.L. acknowledges financial support from the Brazilian agency 
FAPESP (proj. 04/07282-2, proj. 06/57824-1). F.M. acknowledges financial support
from  the Italian  I.N.A.F. (Italian National Institute for
Astrophysics), project PRIN-INAF-2005-1.06.08.16

\end{document}